\documentclass[journal,onecolumn]{IEEEtran}
\usepackage{graphicx}
\usepackage{eurosym}
\usepackage{microtype}
\usepackage{amssymb}
\usepackage{yfonts}
\usepackage{url}
\usepackage{multirow}
\usepackage{xcolor}
\usepackage{blindtext}
\usepackage{eucal}
\usepackage{enumitem}
\usepackage{amsmath}
\usepackage{algorithmic}
\usepackage{nomencl}
\usepackage{etoolbox}
\usepackage{eurosym}
\usepackage{amsfonts}
\usepackage{array}
\usepackage{comment} 
\makenomenclature
\usepackage{booktabs}
\usepackage{cite}
\usepackage{longtable}
\usepackage{threeparttable}
\usepackage{amsfonts} 
\usepackage{times}
\usepackage{nomencl}
\usepackage{subfigure}

\usepackage[caption=false]{subfig}

\makenomenclature

\renewcommand\nomgroup[1]{%
\color{black}
  \item[\bfseries
  \ifstrequal{#1}{P}{Variables}{%
  \ifstrequal{#1}{C}{Constants}{%
  \ifstrequal{#1}{S}{Sets and Indices}{}}}%
]\vspace{7pt}}

\makeatletter
\newcommand{\thickhline}{%
    \noalign {\ifnum 0=`}\fi \hrule height 0.75pt
    \futurelet \reserved@a \@xhline
}
\newcolumntype{"}{@{\hskip\tabcolsep\vrule width 0.75pt\hskip\tabcolsep}}
\makeatother

\makeatletter
\newcommand*{\rom}[1]{\expandafter\@slowromancap\romannumeral #1@}
\makeatother
\usepackage{supertabular}
\usepackage{array}
\usepackage{setspace}
\newenvironment{myenv}[1]
{\begin{spacing}{#1}}
    {\end{spacing}}
\usepackage{anysize}
\marginsize{15mm}{15mm}{15mm}{20mm}

\begin{document}

\title{Controllable Transmission Networks Under Demand Uncertainty with Modular FACTS}
\author{Alireza~Soroudi \textit{}
\thanks{
Alireza Soroudi(alireza.soroudi@ucd.ie) is with the School of Electrical and Electronic Engineering, University College Dublin, Ireland. This work was conducted at the Energy Institute, University College Dublin, Ireland. This publication has emanated from research conducted with financial support from Science Foundation Ireland, Grant Number SFI/15/SPP/E3125.
} 
}

\markboth{}%
{Shell \MakeLowercase{\textit{et al.}}: Bare Demo of IEEEtran.cls for Journals}
\maketitle

\begin{abstract}
The transmission system operators (TSOs) are responsible to provide secure and efficient access to the transmission system for all stakeholders. This task is gradually getting challenging due to the demand growth, demand uncertainty, rapid changes in generation mix, and market policies. Traditionally, the TSOs try to maximize the technical performance of the transmission network via building new overhead lines or physical hardening. However, obtaining public acceptance for building new lines is not an easy step in this procedure. For this reason, the TSOs try to capture the capabilities of existing assets. This paper investigates the use of modular FACTS devices (to alter the line characteristics) for improving the capability of transmission network in serving the uncertain demand without the need for building new overhead lines. The proposed method considers the uncertainty of demands and controls the utilization of existing transmission assets. The mathematical results obtained are validated with a complete non-linear simulation model of three transmission networks.
\end{abstract}
\begin{IEEEkeywords}
Transmission network, Power flow controller, Congestion, Flexibility.
\end{IEEEkeywords}

\section*{NOMENCLATURE}
The notations and symbols used throughout the paper are stated in this section.\\
\begin{myenv}{1.1}
    \begin{supertabular}{>{\arraybackslash}p{1.1cm} >{\arraybackslash}p{18.5cm} }
        \textbf{Abbreviations:} & \\
         $\textit{MF}$& Membership Function\\
        $\textit{M-FACTS}$& Modular Flexible AC Transmission system\\
        $\textit{P2H}$& Power to hydrogen.\\
        $\textit{RES}$& Renewable Energy Sources \\
        $\textit{TSO}$& Transmission System Operator\\ \\
        
        \textbf{Sets:} & \\
        $\Omega_{B}$& Set of network buses.\\
        $\Omega_{\ell}$& Set of network lines.\\
        $\Omega_{G}$& Set of generating units.\\ \\
        \textbf{Variables:}&\\ \\
        $\alpha^f(P)$ & Forecasted membership function describing the uncertain demand \\
        $\alpha^c(P)$ & Calculated membership function describing the uncertain demand considering technical constraints\\
        $LR$ & Load repression \\
        $P_{i,D}^{\alpha}$ & Uncertain active demand at bus i. \\
        $P_{i,g}^{\alpha}$ & Uncertain active output of generator at bus i. \\
        $\delta_{i}^{\alpha}$ & Uncertain delta angle of bus i. \\ \\
        \textbf{Parameters:}&\\ \\
          $\bar{\alpha}^{LR}_{min}$ & Degree of load repression for load reduction \\
        $\bar{\alpha}^{LR}_{max}$ & Degree of load repression for load increase \\
        
        $\beta^{min}_{i,j}$ & Minimum flexibility of M-FATCS for changing the line impedance of the line connecting bus i to j. \\
        $\beta^{max}_{i,j}$ & Minimum flexibility of M-FATCS for changing the line impedance of the line connecting bus i to j. \\
        ${P}_{i,g}^{min}$ & Minimum operating limit of the generator connected to bus i. \\
        ${P}_{i,g}^{max}$ & Maximum operating limit of the generator connected to bus i. \\
           $\bar{P}_{i,D}^{\alpha}$ & Predicted active demand at bus i. \\
        $\check{P}_{i,D}^{\alpha}$ & Predicted lower bound active demand at bus i. \\
        $\hat{P}_{i,D}^{\alpha}$ & Predicted upper bound active demand at bus i. \\
        $B_{i,j}$ & Susceptance of the line connecting bus i to j. \\
        ${P}_{i,j}^{lim}$ & Thermal limit of line connecting bus i to j. \\
        $\bar{\tau}$ & Total available M-FACTS flexibility. \\
        $w_i$ & Weight factor indicating the importance of demand at bus i. \\
        \end{supertabular}
\end{myenv}

\section{Introduction}\label{sec_1}
\subsection{Motivations}
\IEEEPARstart{T}{he} role of transmission system operator is to keep the security and efficiency of the transmission network. The resilient transmission network should be able to deal with high impact low probability events \cite{amraee2017controlled}. The events can be named as natural disasters, cyber-attacks \cite{mohammadpourfard2020ensuring}, physical attacks, simultaneous multiple component outages and etc. Another event that has the potential to cause cascading outages is transmission line overloading. In case, the throughput of a line is beyond the seasonal thermal limits then the protection system might cause an outage of this line and naturally the flow will be transferred to other parallel paths (if they exist). Another solution might be load shedding to alleviate the line's power flow. However, it is not a desirable or cheap solution for the customers.   
The adequacy of the transmission network in supplying the demand can be affected by different factors such as line impedance, thermal capacity, and network topology. Building new overhead lines (to address these problems) usually provokes the protest of residents mainly due to the land-use conflicts, noise, aesthetic concerns, and safety issues \cite{FURBY198819}.\\
\indent TSO should be able to efficiently use the existing assets to ensure the ability of the transmission system to meet reasonable demands \cite{GOUVEIA20091012,sastry2020distributed}. Some important factors should be taken in to account namely, the ideal solution should be rapidly applicable, capable of dealing with uncertainties \cite{5471073}, and finally, community acceptance. The Modular Flexible AC Transmission System (M-FACTS) devices can improve asset utilization and relieving the technical constraints by controlling the power flow in the transmission system \cite{4039419}. 
The M-FACTS \cite{8959319} are modular units that can be installed on some transmission lines and they are capable of pushing the power from heavily loaded lines to lightly loaded lines in the inductive mode of operation. In capacitive mode, the M-FACTS (installed on less loaded lines) will absorb the power from other lines. The M-FACTS devices have some advantages over the FACTS devices such as
fast installation, redeployment possibility, and more flexibility for changing the line impedance \cite{1581595}.  
\subsection{Literature review}
The M-FACTS have been studied in different power system contexts. For example, 
\begin{itemize}
 \item \textcolor{black}{Cyber-Security constrained Placement of FACTS Devices \cite{9113302} based on topological characteristics of the underlying physical graphs of transmission networks.}
    \item Resolving intact and post-contingency overloading \cite{8973921} \textcolor{black}{ using a DC-OPF and Power Transfer Distribution Factors.}
     \item \textcolor{black}{Power quality enhancement in presence of RES technologies \cite{9110836} }
     \item \textcolor{black}{Detecting False Data Injection Attacks on Power Grid State Estimation \cite{8735923}. }
    \item Phase current balancing \cite{kovalsky2020fast}
using the existing logistics infrastructure \cite{kreutzer2020fields}. \textcolor{black}{This industrial project assessed the impact of M-FACTS on the 63-kV grid along with the Wi-Fi communication system between the onsite modules and a 3G network linking the site to a web-hosted platform installed in regional control centers.}
    \item Improving the sustainability of new designs and components installed on the grid \cite{9007818}
    \item Transient stability improvement \cite{tupitsina2020effect} \textcolor{black}{by increasing the fault clearance time and finding the optimal settings of M-FACTS.}
    \item Defending against False Data Injection Attacks \cite{liu2020optimal}.  \textcolor{black}{This reference provided a trade-off between loss minimization and improving the defense strategy.} 
    \item Reliability improvement of power system \cite{anello2017optimal}  \textcolor{black}{using a heuristic method (PSO) linked with industrial package PSSE.}.
    \item Active loss minimization \textcolor{black}{using PSO method.} \cite{radhakrishna2016minimization}
\end{itemize}

Some factors should be considered to decide about the optimal allocation of M-FACTS in transmission network such as:
\begin{itemize}
    \item Mechanical strength of the line which is going to host the M-FACTS. The weight of the modular units should be tolerated by the cables and towers. 
    \item There should be at least one parallel path to push/pull power to/from that path. 
    \item The number of hours as well as the magnitude of risk that can be resolved using M-FACTS should be calculated. For example, if the impact of M-FACTS is reducing 1\% of overload for a limited number of hours (out of 8760 hrs) then it might not be economically viable for investment.
    \item If there is more than one line equipped with M-FACTS then the coordination between their actions is the key element. The central or decentralized control of these devices should be carefully controlled \cite{hala2018unified}.  
    \item The M-FACTS need a minimum current to be able to operate. This is fine with reactive mode (which is helpful for heavily loaded lines) but it might cause problems in the capacitive mode of operation. 
    \item The optimal allocation/operation of M-FACTS is highly dependant on the input assumptions and data of the problem. For example, demand profile, network configuration, RES generation, etc. 
\end{itemize}

\textcolor{black}{Although the placement of Modular FACTS controllers has been widely studied in different research works, no study has considered the possible impacts of these technologies in handling the demand uncertainty and load repression concept.}

\textcolor{black}{
\subsection{Contributions}
The contributions of this work are twofold: 
\begin{itemize}
    \item Optimally allocate and operate M-FACTS devices in the transmission network. This will consider both modes of operation namely, inductive and capacitive modes. Some practical considerations are also taken into account such as eliminating the some lines locations due to practical issues such as (transformers, the poles that are not mechanically strong enough to hold M-FACTS, the lines under seasonal maintenance and etc.). 
    \item Considering the net demand uncertainty using a fuzzy technique.  
\end{itemize}
}

\subsection{Paper structure}
The proposed methodology is described in section \ref{sec_2}. The simulation results are provided and discussed in section \ref{sec:SR}. Finally, the paper is concluded in section \ref{sec:conclusion}. \section{Methodology}
 \label{sec_2}
\begin{figure}[!ht]
  \centering
  \includegraphics[width=0.35\columnwidth,bb=140 360 460 800,angle=0]{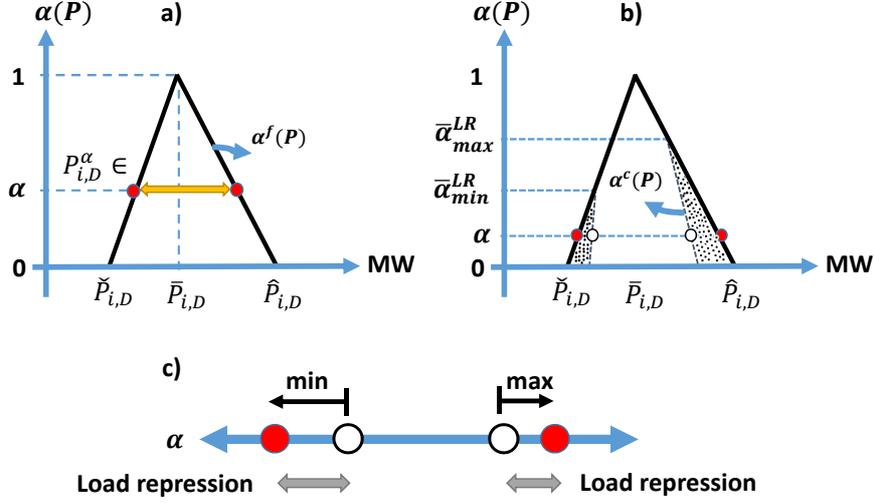}
  	\caption{The load repression concept }\label{fig:LR}
\end{figure}

The objective is to maximize the abbot of the transmission network in supplying the demand even if the precise demand is unknown. There are several techniques to model the demand uncertainty such as scenario based modeling \cite{conejo2010decision} and fuzzy methods \cite{4435948}. The scenario based approach requires the probability density function of the uncertain parameter but the fuzzy approach can be utilised by expert opinion. In this work, the uncertainty of demand is described using a triangle fuzzy membership function (MF) \cite{gouveia2016constrained} as shown in Fig. \ref{fig:LR}-a and mathematically described in (\ref{eq:minalpha})-(\ref{eq:maxalpha}) $\forall i \in \Omega_B$, the forcasted value of $\alpha^f(P)$ is described as: 
\begin{subequations}
\label{eq:alphaf}
\begin{alignat}{2}
\label{eq:minalpha}
& P^{\alpha}_{i,D} \geq \bar{P}_{i,D}-(1-\alpha)(\bar{P}_{i,D}-\check{P}_{i,D}) \\
\label{eq:maxalpha}
& P^{\alpha}_{i,D} \leq \bar{P}_{i,D}-(1-\alpha)(\bar{P}_{i,D}-\hat{P}_{i,D})
\end{alignat}
\end{subequations}

$\bar{P}_{i,D},\hat{P}_{i,D}$ and $\check{P}_{i,D}$ in \eqref{eq:minalpha}, 
\eqref{eq:maxalpha} are the forecasted, upper and lower values for power demand at $\alpha=0$, respectively.
\textcolor{black}{Without loss of generality, the triangle MF is used in this paper. Other fuzzy membership functions can be used for describing the demand uncertainty such as trapezoidal membership functions \cite{914524}.}

The demand should be able to reach the limits specified in \eqref{eq:alphaf} (at every $\alpha$-cut) if the transmission network is fully adequate.  \\
\indent The concept of load repression (LR) was proposed in \cite{matos2008fuzzy}, which uses the $\alpha$-cut technique \cite{mabuchi1988approach} to specify the amount of demand that can not be served in a given bus. The total $LR$ in the system is calculated by calculating the $\alpha^c(P)$ as follows:
\begin{subequations}
\label{eq:baseLR}

\begin{alignat}{3}
\label{eq:obj}
&\alpha^c(P)= \min/\max \sum_i w_i P^\alpha_{i,D} &  \\
\label{eq:balance}
& P^\alpha_{i,g}-P^\alpha_{i,D} = \sum_j(P^\alpha_{i,j}) & \forall i \in \Omega_B  \\
\label{eq:flow}
& P^\alpha_{i,j}=B_{i,j}(\delta^\alpha_i-\delta^\alpha_j) & \forall ij \in \Omega_\ell \\
\label{eq:linelimit}
& -P^{lim}_{i,j} \leq P^\alpha_{i,j} \leq  P^{lim}_{i,j} & \forall ij \in \Omega_\ell\\
\label{eq:genlimit}
& P^{min}_{i,g} \leq P^\alpha_{i,g} \leq P^{max}_{i,g} & \forall i \in \Omega_B, \forall g \in \Omega_G\\
& \text{Constraint} \ \eqref{eq:alphaf} \notag
\end{alignat}
\end{subequations}

The weight factors ($w_i$) in \eqref{eq:obj} show the importance of each demand. These weight factors are not determined by optimization and are specified by the planner.
\textcolor{black}{The nodal power balance is modeled in \eqref{eq:balance}. This constraints shows how the level of generation, demand and power transmission should be matched.}
Constraint \eqref{eq:flow} indicates how line flow from bus $i$ to bus $j$ is calculated \textcolor{black}{at each $\alpha$ level.} 
 
 Constraint \eqref{eq:linelimit} specifies the line flow limits. \textcolor{black}{These limits can be static or dynamic and usually show the thermal limits of the transmission lines at a given time period.} 
 Finally, constraint \eqref{eq:genlimit} ensure the generation limits. \textcolor{black}{It is assumed that the on/off states of the units are predetermined using unit commitment models.}\\
\textcolor{black}{The load repression (LR) has two indices namely, severity and degree. These two concepts are explained here:\\}
The \textit{severity} of $LR$ is shown in Fig.\ref{fig:LR}-b (dotted area). The \textit{degree} of $LR$ is the maximum value of $\alpha$ where $LR$ does not happen. This quantity is depicted in Fig.\ref{fig:LR}-b. The degree of load repression might be different for load reduction ($\bar{\alpha}^{LR}_{min}$) and load increase ($\bar{\alpha}^{LR}_{max}$).
Fig.\ref{fig:LR}-c indicates how $LR$ is calculated at each $\alpha$-level in  maximization/minimization of demand for a given bus $i$. Fig.\ref{fig:LR}-c also demonstrates why two separate optimization (min and max) are needed in \eqref{eq:baseLR}. \\
Once \eqref{eq:baseLR} is solved then the total $LR$ is calculated as follows:
\begin{alignat}{2}
\label{eq:LRcalc}
LR=\int_{\check{P}_{i,D}}^{\hat{P}_{i,D}} (\alpha^f(P)-\alpha^c(P) ) dP_{i,D}
\end{alignat}

$\alpha^f(P)$ and $\alpha^c(P)$ are the forecasted MF \eqref{eq:alphaf} and the calculated MF (\eqref{eq:baseLR} or \eqref{eq:LRFACTS}), respectively. 
\textcolor{black}{The forecasted values of demand are obtained from historic data or expert's opinion.}
By deploying a M-FACTS device in a given line, the line's reactance becomes a decision variable and may contribute to $LR$ reduction if it is optimally \textcolor{black}{ allocated and operated}. 
The mathematical formulation for $LR$ calculation in presence of M-FACTS is as follows: 
\begin{subequations}
\label{eq:LRFACTS}
\begin{alignat}{3}
& \alpha^c(P)=\min/\max \sum_i w_i P^\alpha_{i,D} & \\
& P^\alpha_{i,g}-P^\alpha_{i,D} = \sum_j(P^\alpha_{i,j}) & \forall i \in \Omega_B  \\
\label{eq:flow2A}
& P^\alpha_{i,j}=(B_{i,j})(1+\beta_{i,j})(\delta^\alpha_i-\delta^\alpha_j) & \forall ij \in \Omega_\ell \\
& \beta^{min}_{i,j} \leq \beta_{i,j} \leq  \beta^{max}_{i,j} & \forall ij \in \Omega_\ell \\
& -P^{lim}_{i,j} \leq P^\alpha_{i,j} \leq  P^{lim}_{i,j} & \forall ij \in \Omega_\ell\\
& P^{min}_{i,g} \leq P^\alpha_{i,g} \leq P^{max}_{i,g} & \forall i \in \Omega_B, \forall g \in \Omega_G\\
& \text{Constraint} \ \eqref{eq:alphaf} \notag
\end{alignat}
\end{subequations}

The value of $\beta_{i,j}$  in \eqref{eq:flow2A} shows that how much change in the susceptance of the line between bus $i$ and $j$ ($B_{i,j}$) is going to happen (M-FACTS flexibility). The reactance variation is limited to the characteristics of the M-FACTS device ($\beta^{min/max}_{i,j}$). If it is pure capacitive then $\beta^{min}_{i,j}=0$ and $\beta^{max}_{i,j}\geq 0$. Alternatively, if it is working in inductive mode then $\beta^{min}_{i,j}\leq 0$ and $\beta^{max}_{i,j}= 0$.\\
The M-FACTS deployed on a specific transmission line can be operated in different modes based on the operating condition of the system. \textcolor{black}{For example, the heavily loaded lines are the best locations for M-FACTS in inductive mode of operation. Alternatively, the lightly loaded lines are the best locations for M-FACTS in capacitive mode of operation}.\\
In reality, the TSO has a limited amount of M-FACTS devices available \textcolor{black}{to deploy in the system}  so it is essential to efficiently utilize them in normal and contingency conditions. The following optimization answers this question:
\begin{subequations}
\label{eq:FACTSallocation}
\begin{alignat}{2}
&\alpha^c(P)= \min/\max \sum_i w_i P^\alpha_{i,D} \\
& \sum_{i,j \in \Omega_\ell}  | \beta_{i,j}| \leq \bar{\tau} \\
& P^\alpha_{i,g}-P^\alpha_{i,D} = \sum_j(P^\alpha_{i,j}) & \forall i \in \Omega_B  \\
& P^\alpha_{i,j}=(B_{i,j})(1+\beta_{i,j})(\delta^\alpha_i-\delta^\alpha_j) & \forall ij \in \Omega_\ell \\
& \beta^{min}_{i,j} \leq \beta_{i,j} \leq  \beta^{max}_{i,j} & \forall ij \in \Omega_\ell \\
& -P^{lim}_{i,j} \leq P^\alpha_{i,j} \leq  P^{lim}_{i,j} & \forall ij \in \Omega_\ell\\
& P^{min}_{i,g} \leq P^\alpha_{i,g} \leq P^{max}_{i,g} & \forall i \in \Omega_B, \forall g \in \Omega_G\\
& \text{Constraint} \ \eqref{eq:alphaf} \notag
\end{alignat}
\end{subequations}
where $\bar{\tau}$ is the total available M-FACTS flexibility in the system. 

\begin{figure}[!ht]
  \centering
	\includegraphics[width=0.5\columnwidth,angle=90]{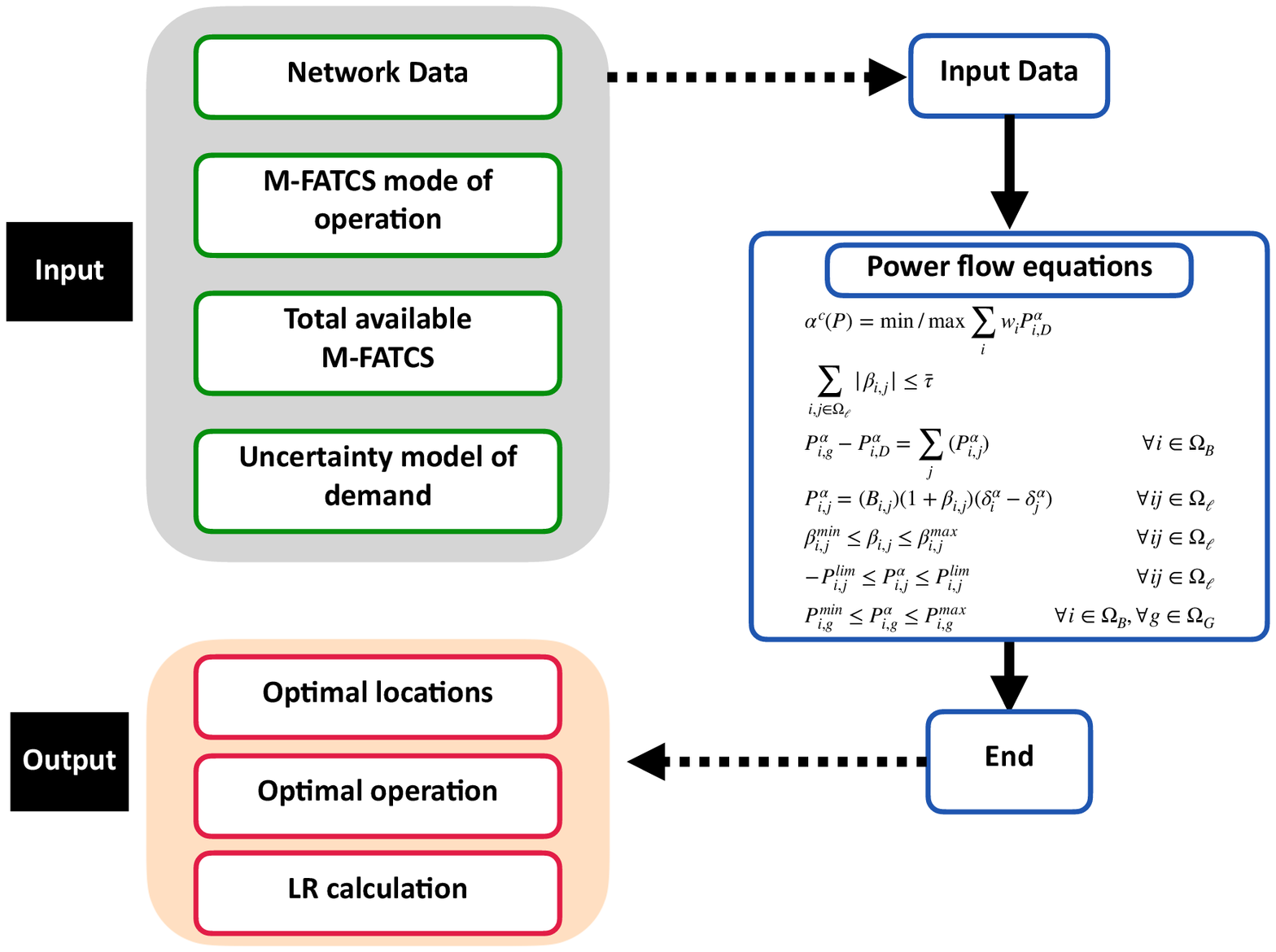}
	\vspace{0mm}
  \caption{The overall structure of the proposed model for M-FACTS allocation}
\label{fig:chart}
\end{figure}

The overall structure of the proposed model for M-FACTS allocation is shown in Fig. \ref{fig:chart}.\\
\vspace{10mm}

\section{Simulation Results}
\label{sec:SR}
The proposed framework in \eqref{eq:LRFACTS} and \eqref{eq:FACTSallocation}
are implemented in GAMS \cite{Soroudi2017} and the non-linear problem is solved using KNITRO solver \cite{waltz2004knitro}. The simulations are implemented on two transmission networks namely 5-bus PJM network, IEEE 24-bus system and \textcolor{black}{IEEE 118-bus system}.  

\subsection{Five bus system}
\label{ssec:case5bus}
\begin{figure}[!ht]
  \centering
	\includegraphics[width=0.4\columnwidth,angle=-90]{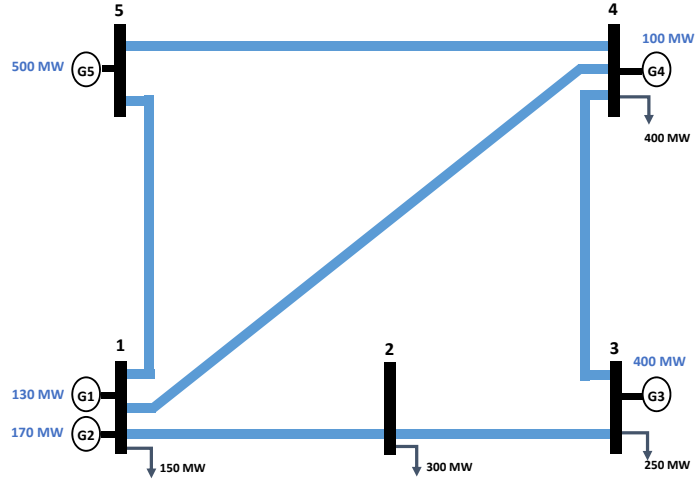}
  \caption{Five bus network for explaining the repression concept}
\label{fig:bus5}
\end{figure}

The data of the system under study is shown in 
Fig. \ref{fig:bus5}. \textcolor{black}{This network has 6 transmission lines 4 load points. }
\vspace{-8mm}
The network characteristic data is provided in Table \ref{tab:data5bus}. 

\begin{table}[ht]
\centering
\caption{Line characteristics of 5 bus network}
\label{tab:data5bus}
  \begin{tabular}{cccc}
\hline
from bus	&	to bus	&	X (PU)	&	Limit (MW)	\\\hline
1	&	2	&	0.030	&	240	\\
1	&	4	&	0.050	&	270	\\
1	&	5	&	0.060	&	250	\\
2	&	3	&	0.025	&	270	\\
3	&	4	&	0.030	&	270	\\
4	&	5	&	0.020	&	270	\\ \hline 
  \end{tabular}
\end{table}

The forecasted values of demand ($\bar{P}_{i,D}$) are specified in Fig. \ref{fig:bus5}. The upper and lower values of demand ($\hat{P}_{i,D}, \check{P}_{i,D}$) are assumed to be $\pm5\%$ more/less than the forecasted values. The forecast error can be different from what is considered in this work but the general concept of the proposed framework remains valid. 

\begin{figure}[!ht]
  \centering
  \includegraphics[width=0.5\columnwidth,angle=0]{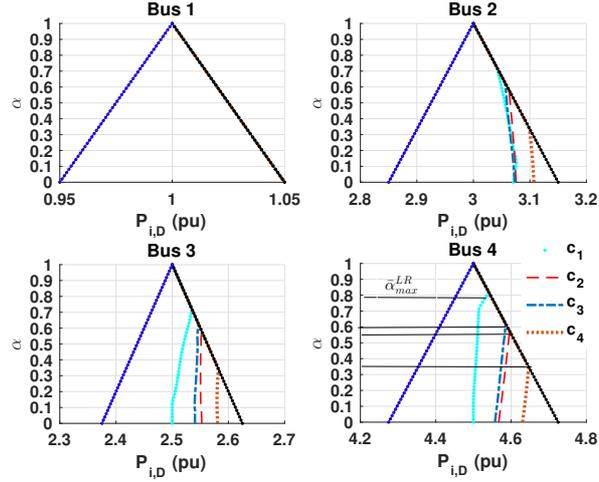}
  \vspace{-4mm}
  	\caption{The $LR$ for $c_1$: Base, $c_2$: Inductive, $c_3$: Capacitive, $c_4$: Smart in five bus network case }
  \label{fig:LR5bus}
\end{figure}

The $LR$ calculation is done for four different strategies namely:
\begin{itemize}
    \item Strategy $c_1$: Base case in which no M-FACTS device exists in the network $\beta^{min/max}_{i,j}=0$. The network impedances are unchanged in this strategy. 
    \item Strategy $c_2$: Inductive M-FACTS  $\beta^{min}_{i,j}\leq 0$ , $\beta^{max}_{i,j}=0$. In this case, the reactance increase is used for reducing the line loading. 
    \item Strategy $c_3$: Capacitive M-FACTS $\beta^{min}_{i,j}=0$ , $\beta^{max}_{i,j}\geq 0$. In this case, the reactance reduction is used for attracting flow from the heavily loaded lines. 
    \item Strategy $c_4$: Smart (coordinated Inductive-Capacitive) M-FACTS are considered
    $\beta^{min}_{i,j}\leq 0$ , $\beta^{max}_{i,j}\geq 0$. The reactance of some lines are increase while reducing the reactance on some other lines. 
\end{itemize}
It is assumed that all lines are equipped with M-FACTS devices ($c_{2-4}$) and all $w_i$ in \eqref{eq:LRFACTS} are equal. The M-FACTS capacity is assumed to be a percentage of the line's impedance. It is assumed that the actions of all M-FACTS are coordinated with each other. 
The load repression in all buses with demand are depicted for different control strategies in Fig.\ref{fig:LR5bus}. 
\textcolor{black}{Fig.\ref{fig:LR5bus} is explained as follows: Bus 1 does not show any load repression. The reason is that this bus has sufficient generation capacity and is not dependant on the transmission system adequacy. The bus 5 is not depicted since it does not have any demand connected to it. The remaining buses (bus 2,3,4) show load repression (in load increase direction). The degree and severity of load repression in $c_4$ case (both inductive and capacitive) model of operation is less than three other cases. The base case $c_1$ (with no M-FACTS), the degree and severity of LR is maximum.
}\\
The total $LR$ is calculated using \eqref{eq:LRcalc} and the results are given in table \ref{tab:case5bus}. It can be observed that not only the total $LR$ reduces in $c_4$ (compared to $c_1$) but also the degree of repression ($\bar{\alpha}^{LR}_{max}$) decreases. 
The optimal dispatch ($\beta_{i,j}$) decision for M-FACTS devices in different strategies are given in table \ref{tab:case5busdsr}.

\begin{table}[ht]
\centering
\caption{Total $LR$ for 5 bus system in different strategies}
\label{tab:case5bus}
  \begin{tabular}{ccccc}
\hline
Strategy & $c_1$  & $c_2$  & $c_3$ & $c_4$ \\ \hline
LR (MW) & 17.427 & 8.926 &  10.396  & 3.255 \\ \hline
$\bar{\alpha}^{LR}_{i=1,max}$ &  0&0&0& 0\\ \hline 
$\bar{\alpha}^{LR}_{i=2,max}$ &  0.70&0.57&0.62& 0.33\\ \hline 
$\bar{\alpha}^{LR}_{i=3,max}$ &  0.70&0.58&0.61& 0.32\\ \hline 
$\bar{\alpha}^{LR}_{i=4,max}$ &  0.80&0.55&0.60& 0.35\\ \hline 
  \end{tabular}
\end{table}
\begin{table}[ht]
\centering
\caption{M-FACTS dispatch ($\beta_{i,j}$) for 5 bus system in different strategies}
\label{tab:case5busdsr}
  \begin{tabular}{c|c|c|c}
\hline
FACTS location ($i-j$) &  $c_2$  & $c_3$ & $c_4$ \\ \hline 

2-1	&	-0.1499	&	0.0520	&	-0.0965	\\
4-1	&		&	0.2000	&	0.2000	\\
5-1	&		&	0.2000	&	0.2000	\\
3-2	&	-0.0760	&	0.1196	&	0.1171	\\
4-3	&	-0.1057	&	0.1054	&	-0.0561	\\
5-4	&	-0.2000	&	    	&	-0.2000	\\  \hline
  \end{tabular}

\end{table}

\begin{figure}[!htb]
  \centering
  \includegraphics[width=0.5\columnwidth,angle=0]{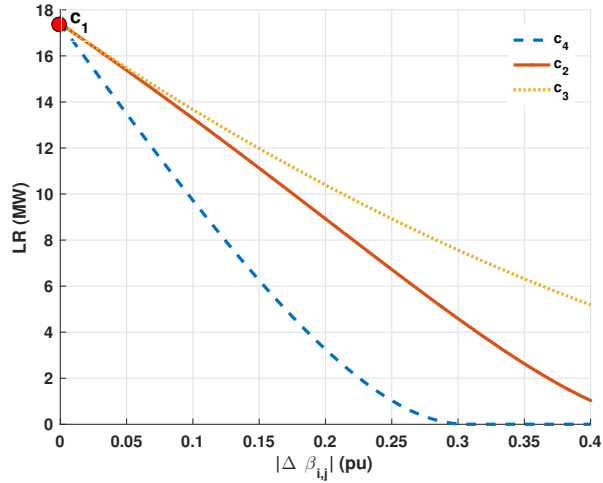}
  	\caption{Impact of M-FACTS capacity on load repression of 5 bus network in different strategies: Sensitivity analysis; $c_2$: Inductive , $c_3$: Capacitive , $c_4$: Smart}\label{fig:sen}
\end{figure}
The capacity of M-FACTS devices are changed from 0 to 40\% and the impact on total $LR$ are shown in Fig. \ref{fig:sen}. The most effective strategy is inductive-capacitive coordination ($c_4$) for this system. 

\subsection{IEEE 24 bus system}
\label{ssec:case24bus}
\begin{figure}[!ht]
  \centering
\includegraphics[width=0.5\columnwidth,angle=0]{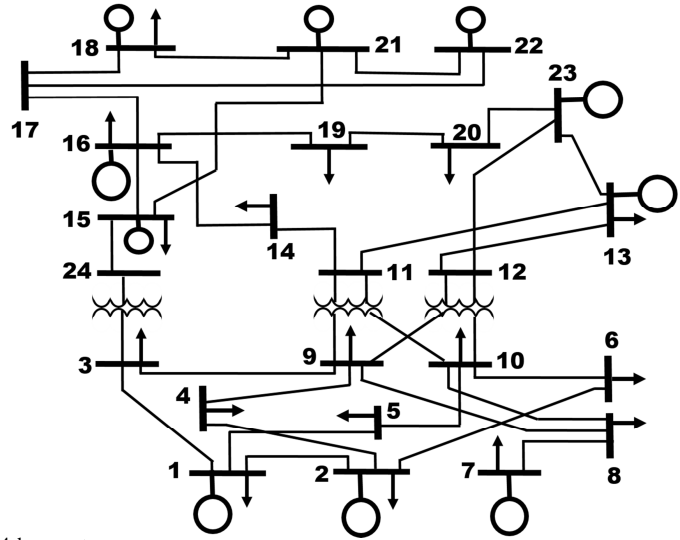}
  \vspace{-40mm}
  	\caption{Single line diagram of IEEE 24 bus system}\label{fig:ieee24bus}
\end{figure}

\begin{figure}[!ht]
  \centering
  \includegraphics[width=0.5\columnwidth,angle=0]{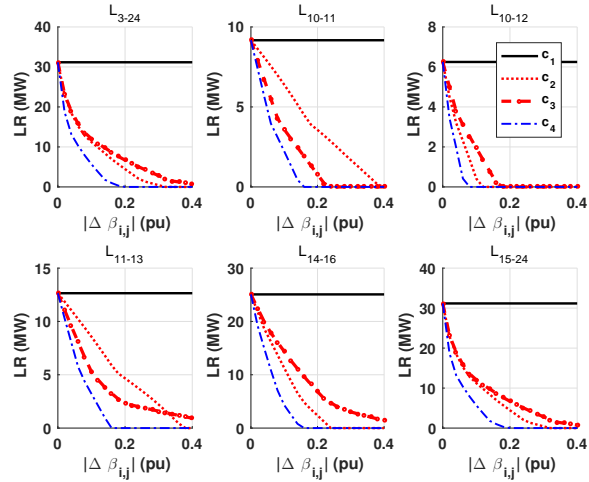}
  	\caption{Impact of M-FACTS capacity and line contingencies on the load repression of IEEE 24 bus network in different strategies: Sensitivity analysis; $c_2$: Inductive, $c_3$: Capacitive, $c_4$: Smart}\label{fig:LR24bus}
\end{figure}

The single line diagram of IEEE 24 bus system is depicted in Fig. \ref{fig:ieee24bus}. The data of this system is available in \cite{7958953}. In this case, it is assumed that $\check{P}_{i,D}=0.9\bar{P}_{i,D}$ and $\hat{P}_{i,D}=1.1\bar{P}_{i,D}$. This system does not show any load repression in the intact condition. 
Now, the impact of all single-line outages on $LR$ are investigated and the results are shown in Fig. \ref{fig:LR24bus}.
\textcolor{black}{The most severe contingencies are the outages of $L_{3-24} \& L_{15-24}$ where $LR=30MW$ in base case ($c_1$). }

\begin{figure}[!ht]
  \centering
  \includegraphics[width=0.5\columnwidth,angle=0]{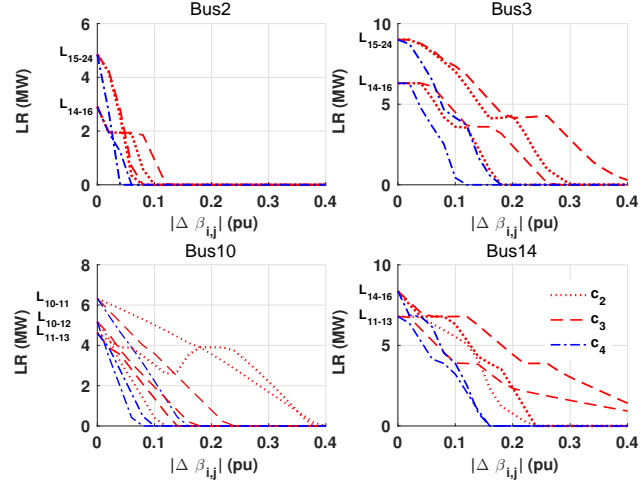}
  	\caption{Impact of contingencies and M-FACTS capacity on load repression of different buses in each strategy; $c_2$: Inductive , $c_3$: Capacitive , $c_4$: Smart in IEEE 24 bus system}\label{fig:busLR}
\end{figure}

Similar to the previous case, the capacity of M-FACTS devices are changed from 0 to 40\% and the impact on total $LR$ are shown in Fig. \ref{fig:busLR} and the line contingencies with $LR\geq 0$ are plotted. As it can be observed in this figure, the most effective strategy is inductive-capacitive coordination ($c_4$) for this system.

\begin{figure}[!ht]
  \centering
  \includegraphics[width=0.55\columnwidth,angle=0]{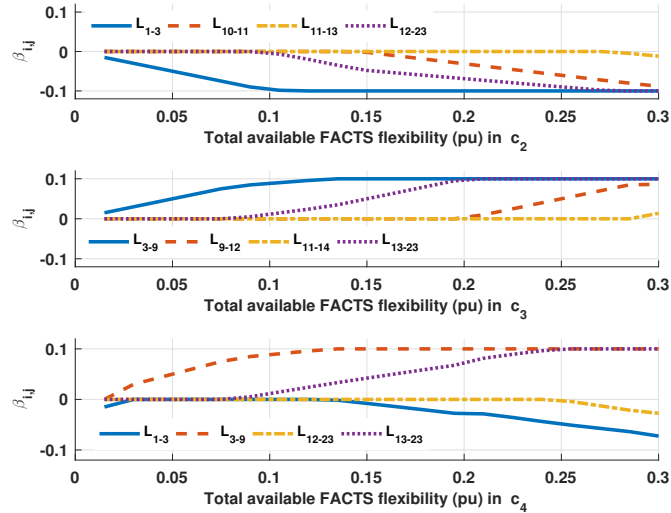}
  \vspace{0mm}
  	\caption{The optimal M-FACTS deployment vs available M-FACTS flexibility in each strategy when $L_{15-24}$ is out of service in IEEE 24 bus system} \label{fig:strategy}
\end{figure}

It is important to know which buses are contributing to the $LR$ when a contingency happens. The impacts of different contingencies on the load repression vs M-FACTS capacity are shown in Fig. \ref{fig:busLR}. As it is expected, the location of the bus, the amount of load on that bus, and also the contingency affect the $LR$ in each strategy. For example, on bus 3, the $LR$ is 8.99 MW when the line connecting bus 15-24 is out of service. With the increase of M-FACTS capacity, this $LR$ reduces in all strategies ($c_{2,3,4}$). Obviously, the most efficient way of dealing with $LR$ is $c_4$ strategy in which, the $LR$ becomes zero at $|\beta_{i,j}|=0.18$ pu.
The optimal M-FACT deployment/exploitation vs the $\bar{\tau}$ \eqref{eq:FACTSallocation} is plotted in Fig.\ref{fig:strategy} in different strategies. This graph shows how TSO should react against the $L_{15-24}$ outage in different strategies. 
For example, in pure inductive M-FACTS ($c_2$), the order of activating the M-FACTS is $L_{1-3} \xrightarrow{}L_{12-23}\xrightarrow{}L_{10-11}\xrightarrow{}L_{11-13}$. 
\subsection{IEEE 118 bus system}
\label{ssec:case118bus}
\begin{figure}[!ht]
  \centering
	\includegraphics[width=0.7\columnwidth,angle=90]{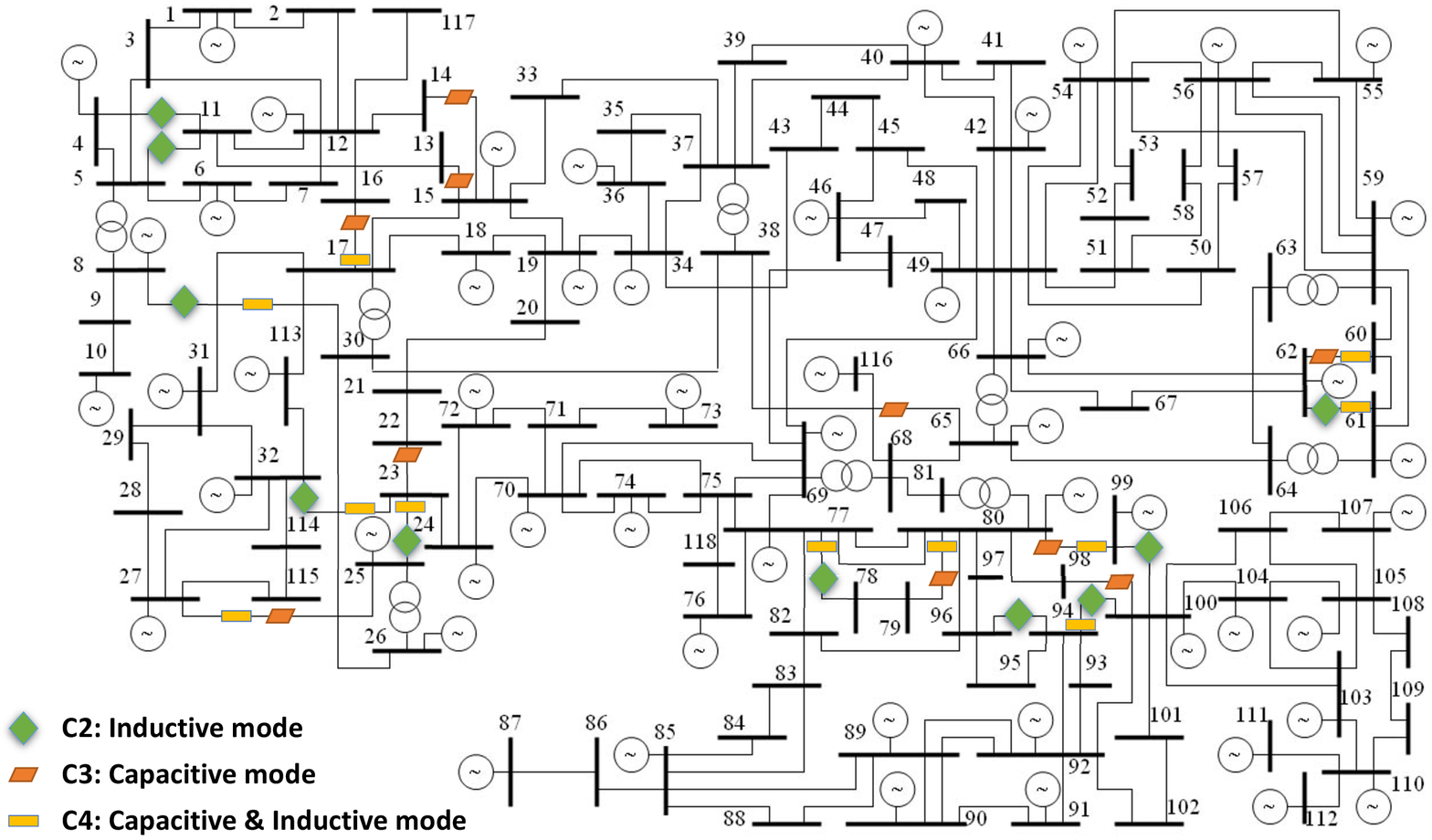}
	\vspace{-20mm}
  \caption{IEEE 118 bus network M-FACTS allocation in different modes of operation}
\label{fig:ieee118bus}
\end{figure}

\textcolor{black}{The proposed model described in \eqref{eq:FACTSallocation}
 is applied to the IEEE 118 bus system. The system data is available in \cite{8119553}. The original system does not show any load repression. In order to put some stress on the system, the thermal limits of the transmission lines are intentionally reduced by 30\%. The $\beta^{max}_{i,j}=\beta^{min}_{i,j}$ is assumed to be 0.15 . The up/down flexibility may not be equal in practice but it does not change the problem formulation. 
 The single line diagram of IEEE 118 bus system as well as the M-FACTS allocation in different strategies ($c_{1,2,3,4}$) is depicted in Fig. \ref{fig:ieee118bus}.}
\textcolor{black}{The total number of candidate lines for hosting the M-FATCS is limited to 12 lines.   
The total load repression in different operating modes are given in Table \ref{tab:case118bus}. }
\begin{table}[ht]
\centering
\caption{\textcolor{black}{Total $LR$ for IEEE 118-bus system in different strategies}}
\label{tab:case118bus}
  \begin{tabular}{ccccc}
\hline
Strategy & $c_1$  & $c_2$  & $c_3$ & $c_4$ \\ \hline
LR (MW) & 76.389 & 51.799 &  53.253  & 45.255 \\ \hline
\hline 
  \end{tabular}
\end{table}
\textcolor{black}{The values of LR in Table \ref{tab:case118bus}, shows a clear reduction when moving from $c_1$ (no M-FACTS) to $c_4$ strategy.
From $LR$ point of view the $c_4$ strategy is the most efficient one for reducing the system's load repression. However, the $c_4$ mode of operation requires more coordination between the M-FACTS. Some of them should work in inductive mode while the rest of them are in capacitive mode. This implies the necessity of communication systems to maintain this coordination.}

\subsection{Future work}
There are several routes that the proposed framework can be extended as follows: 
\begin{itemize}
    \item The impact of flexible transmission network on unit commitment formulation should be investigated. 
    \item A more detailed multi-period AC-OPF can better characterize the impact of M-FACTS on transmission systems. 
    \item The impact of M-FATCS on voltage stability should be investigated. Altering the line impedance will change the load-ability margin of a given system. This should be taken into account to avoid technical problems for the system hosting M-FACTS.  
    \item The uncertainty of wind power generation should be taken into account to avoid financial and technical risks \cite{soroudi2013decision}. \textcolor{black}{If wind uncertainty is modeled using stochastic models then a possibilistic-scenario based model \cite{6142135} can be used.}
    \item The risk of cyber attacks on M-FACTS should be investigated. The attackers may endanger the security of the power system by sending false commands to these devices and cause cascading failures. 
    \item It should be noted that if the power flow is pushed from a heavily loaded line to less loaded line the destination line also remains in safe operating condition. 
    \item Cost-benefit analysis can justify the optimal investment decisions for the TSO. 
\end{itemize}

\section{Conclusion}
\label{sec:conclusion}
The proposed method can provide useful indicators to TSO in effectively dispatching the transmission network using M-FACTS.
The main findings of this paper are outlined as follows:
\begin{itemize}
    \item The optimal operating settings of M-FACTS are should be dynamically updated to cope with the possible changes. The modular configuration of M-FACTS allows them to be redeployed with a short lead time if needed. 
    \item \textcolor{black}{The deployment line of M-FACTS as well as the size of it have significant impacts on the control-ability of the transmission network. This is due to the physical rules behind the OPF equations. It is important to optimally determine the capacity as well as the location of them otherwise, the anticipated flexibility will not be achieved.} 
    \item The optimal deployment strategy will remain valid as far as the network topology remains intact. If it is changed due to maintenance or contingencies then the flexibility of the configured M-FACTs is affected. 
    \item The presence of uncertain renewable power generation can increase the uncertainty of the net load at each given bus. This has the potential to increase the load repression and M-FACTS can be helpful to manage it. 
    \item The M-FACTS is considered as a non-wire solution since it does not require building new overhead lines. However, if the planning of M-FACTS as well as the traditional overhead lines are done simultaneously then a higher level of flexibility will be achieved.  
\end{itemize}

\section*{Acknowledgements}
The work done by Alireza Soroudi is supported by a research grant from Science Foundation Ireland (SFI) under the SFI Strategic Partnership Programme Grant No. SFI/15/SPP/E3125. The opinions, findings and conclusions or recommendations expressed in this material are those of the author(s) and do not necessarily reflect the views of the Science Foundation Ireland. 
\bibliographystyle{IEEEtran}
\bibliography{ref}

\end{document}